\documentclass[aps,twocolumn,prl,amsmath,amssymb,amsfonts]{revtex4-1} 
\usepackage[T1]{fontenc}
\usepackage{mathptmx}
\usepackage[latin9]{inputenc}
\usepackage{verbatim}
\usepackage{textcomp}
\usepackage{amsmath}
\usepackage{amssymb}
\usepackage{graphicx}
\usepackage{color}
\usepackage[english]{babel}
\usepackage{booktabs}
\begin{document}

\title{Relativistic Measurement Backaction in the Quantum Dirac Oscillator}

\author{Keye Zhang$^{1,4}$}
\email[Email: ]{kyzhang@phy.ecnu.edu.cn}
\author{Lu Zhou$^{1,4}$}
\author{Pierre Meystre$^{2}$}
\author{Weiping Zhang$^{3,4}$}
\email[Email: ]{wpz@sjtu.edu.cn}
\affiliation{$^1$Quantum Institute for Light and Atoms, School of Physics and Material Science, East China Normal University, Shanghai 200241, P.R. China}
\affiliation{$^2$Department of Physics and College of Optical Sciences, University of Arizona, Tucson, AZ 85721, USA.}
\affiliation{$^3$Department of Physics and Astronomy, Shanghai Jiao Tong University, and Tsung-Dao Lee Institute, Shanghai 200240, China}
\affiliation{$^4$Collaborative Innovation Center of Extreme Optics, Shanxi University, Taiyuan, Shanxi 030006, P. R. China}

\begin{abstract}
 
An elegant method to circumvent quantum measurement backaction is the use of quantum mechanics free subsystems (QMFS), with one approach involving the use of two oscillators with effective masses of opposite signs.  Since negative energies, and hence masses, are a characteristic of relativistic systems a natural question is to what extent QMFS can be realized in this context. Using the example of a one-dimensional Dirac oscillator we investigate conditions under which this can be achieved, and identify Zitterbewegung or virtual pair creation as the physical mechanism that fundamentally limits the feasibility of the scheme. We propose a table-top implementation of a Dirac oscillator system based on a spin-orbit coupled ultracold atomic sample that allows for a direct observation of the corresponding analog of virtual pair creation on quantum measurement backaction. 
\end{abstract}

\maketitle

{\it Introduction} -- A major challenge of quantum metrology is the need to minimize the backaction noise that accompanies the measurement of quantum observables. Efforts at circumventing this difficulty have led  to the development of quantum non-demolition measurements, and to the use of nonclassical field states that locate quantum fluctuations where they do not significantly perturb the measurement. More recently, Tsang and Caves~\cite{Tsang2012} and Polzik and coworkers~\cite{Polzik2009} realized that it is sometimes possible to isolate quantum-mechanics free subsystems (QMFS) of a quantum system whose observables are by construction quantum non-demolition observables. 

An example considered in Ref.~\cite{Tsang2012} consists of two harmonic oscillators of identical frequencies and opposite masses described by the Hamiltonian
\begin{equation}
H= \frac{\hat{p}^2}{2m} + \frac12 m\omega^2 \hat x^2 - \frac{\hat{p}'^2}{2m} - \frac12 m\omega^2 \hat x'^2.
\label{eq:2ho}
\end{equation}
Considering the composite variables $\hat X= \hat x+\hat x'$, $\hat{P}=\frac12(\hat{p}+\hat{p}')$, $\hat{\Phi}=\frac12(\hat x-\hat x')$ and $\hat{\Pi}=\hat{p}-\hat{p}'$, with $[\hat X, \hat \Pi]= [\hat P, \hat \Phi]=0$, it is easily verified that the Heisenberg equations of motion for $\hat X$ and $\hat \Pi$ form the closed system
\begin{equation}
\dot {\hat X}(t)= \frac{\hat{\Pi}(t)}{m} \,\,\,\,\,,\,\,\,\,\dot{ \hat{\Pi} }= -m\omega^2 \hat X(t),
\label{Hs1}
\end{equation}
so that~$\hat X$ and $\hat \Pi$, the collective position and relative momentum, form a QMFS that allows for their simultaneous and repeated measurement with arbitrary accuracy -- and likewise for the pair $\{\hat{\Phi}, \hat{P}\}$. QMFS implementations have been realized in atomic spin ensembles~\cite{Polzik2010}, hybrid optomechanical systems~\cite{Polzik2017}, microwave-coupled mechanical oscillators~\cite{Korppi2016}, and have  been proposed in Bose-Einstein condensates with negative effective mass component~\cite{Zhang2013}.

The fact that the QMFS of Ref.~\cite{Tsang2012} relies on the use of a negative mass oscillator leads one to ask to what extent the negative energy states in relativistic quantum systems can result in the existence of QMFS in these systems as well. This is the question that we address, using the example of a one-dimensional Dirac oscillator. We find that already in that implementation it is fundamentally different from two independent harmonic oscillators of opposite masses, due to the presence of a `spin-orbit' coupling-like term associated with relativistic Zitterbewegung: even in the non-relativistic limit its remnants limit the ability to realize back-action evading measurements -- this is in addition to the known fact that the localization of particles is limited by the Compton wavelength $\lambda_c$. We then quantify the impact of Zitterbewegung in the full relativistic regime. We conclude by proposing a table-top atomic, molecular, and optical physics implementation that permits to demonstrate this behavior in a non-relativistic system.

{\it Model} -- The Dirac oscillator is an extension of the Dirac equation for a free particle that is linear both in position and momentum. It was introduced by Moshinsky {\it et al} \cite{Moshinsky1989}, who added the linear vector potential $-i  {\boldsymbol \beta} m \omega {\hat x}$  to the Dirac equation. In addition to its use in nuclear physics and relativistic quantum physics, see e.g. the reviews~\cite{Sadurni2011, Quesne2017}, it has found applications in fields ranging from condensed matter physics to quantum optics~\cite{Strange2013, Bermudez2008, Bermudez2007,Rozmejdag1999}.

We concentrate on one spatial dimension, in which case the Dirac matrices $\boldsymbol \alpha$ and $\boldsymbol \beta$ reduce to the Pauli operators $\hat{\sigma}_x$ and $\hat{\sigma}_z$, leading to the reduced equation
\begin{equation}
i\hbar\partial_{t}|\Psi\rangle= \left [c\hat{\sigma}_x \hat p -mc\omega\hat{\sigma}_y \hat x +mc^2\hat \sigma_z \right ] |\Psi \rangle \equiv H_{\rm DO} |\Psi \rangle.
\label{HDO}
\end{equation}
The energy spectrum of $H_{\rm DO}$ comprises a positive energy branch with eigenenergies $E_n^+=mc^{2}\sqrt{1+2n\hbar\omega/mc^{2}} \label{En+}$ bounded from below by  $mc^2$ and a negative energy branch with energies $E_n^-=-E_{n+1}^+$ bounded from above by $-mc^2\sqrt{1 + 2 \hbar\omega/mc^2}$,  see Fig.~1(a).

The corresponding eigenstates are $\left|E_n^+\right\rangle  = A_n |n, \uparrow \rangle - i B_n|n-1, \downarrow \rangle$ and $\left|E_n^-\right\rangle  = B_{n+1}|n+1, \uparrow\rangle + i A_{n+1}|n, \downarrow \rangle$, where $|n\rangle$ are the eigenstates of a non-relativistic harmonic oscillator, $\left|\uparrow,\downarrow\right\rangle $ are Pauli spinors and $A_n=\sqrt{(E_n^++mc^2)/2E_n^+}$ and $B_n=\sqrt{(E_n^+-mc^2)/2E_n^+}$.
The fact that the eigenstates of the Dirac oscillator are linear superpositions of the motional and spin states $|n, \uparrow\rangle $ and $|n-1, \downarrow\rangle$ is a consequence of the spin-orbit coupling interaction in $H_{\rm DO}$, a relativistic effect resulting from enforcing first-order spatial dependences in its wave equation. 

Figure~1(b) plots $|A_n|^2$ and $|B_n|^2$ for $n=0 \ldots 3$ as a function of the relativistic parameter $\epsilon \equiv \hbar\omega/mc^2$. For $\epsilon \gg 1$, we have that  $|A_n| \approx |B_n| \approx 1/\sqrt{2}$,  and the eigenstates $|E_n^\pm \rangle$ exhibit spin-orbit coupling and entanglement between motional and spin degree of freedom. In the non-relativistic limit $\epsilon \rightarrow 0$, in contrast, $|A_n| \rightarrow 1$ and $|B_n| \rightarrow 0$, and the eigenstates and eigenenergies reduce to those of two harmonic oscillators of frequency $\omega$ associated with the spin-up and spin-down components. The first one has a positive mass $m$ and ground state energy $mc^2$, and the second one a negative mass $-m$ and energy bound from above by $-(\hbar \omega + mc^2)$.  

{\it Non-relativistic limit} -- For $\epsilon \rightarrow 0$ the Dirac oscillator can, therefore, be approximated by the Hamiltonian
\begin{equation}
H_{\rm nr}=\left (mc^2+\frac{\hat p^2}{2m}+\frac{m\omega^2\hat x^2}{2}\right )\hat\sigma_z-\frac{\hbar\omega}{2},
\label{eq:nrDirac}
\end{equation}
which describes a pair of one-dimensional harmonic oscillators of spin-dependent mass -- positive mass for spin up and negative mass for spin down. While this suggests that it might be possible to form a QMFS similar to that of Ref.~\cite{Tsang2012} there is an important difference between the Hamiltonians~(\ref{eq:nrDirac}) and (\ref{eq:2ho}), the seemingly inconsequential spin operator $\hat \sigma_z$, the remnant of the spin-orbit coupling. We see shortly that it results in profound differences in the dynamics of the two systems.

\begin{figure}
\begin{centering}
\includegraphics[scale=0.45]{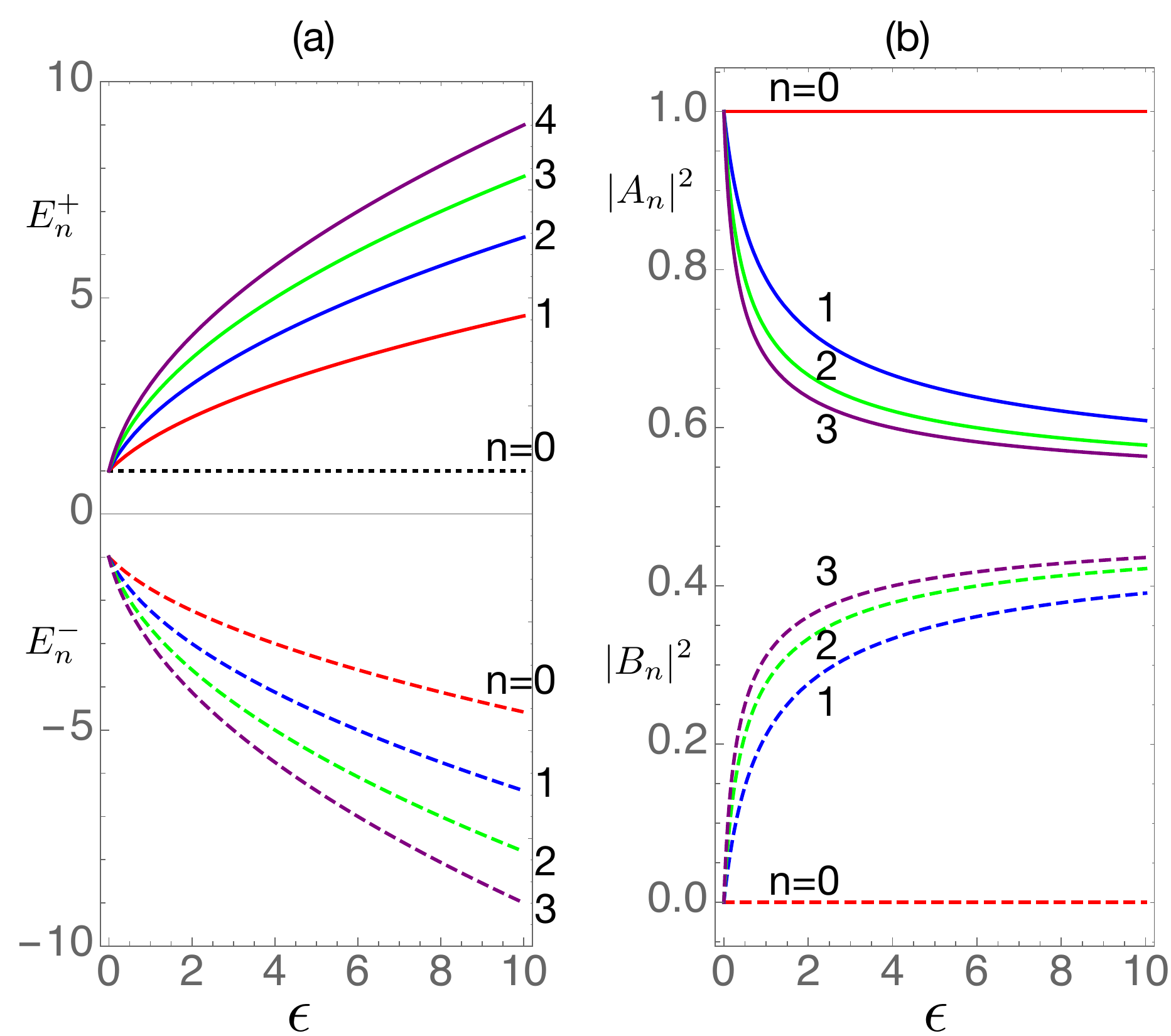}
\par\end{centering}
\caption{(a) Eigenenergies $E_n^{\pm}$, $n=0, \ldots 4$, in units of $mc^2$ as a function of the ratio $\epsilon = \hbar \omega/mc^2$. (b) Probabilities $|A_{n}|^{2}$ and $|B_{n}|^{2}$, $n=0 \ldots 3$, of the two components of the eigenstates.}
\end{figure}

In analogy with Ref.~\cite{Tsang2012} we introduce the operators $\hat X = \hat x \boldsymbol{ \it I}=\hat x |\uparrow \rangle \langle \uparrow |+\hat x |\downarrow \rangle \langle \downarrow |\equiv \hat x_\uparrow + \hat x_\downarrow$ and $\hat \Pi = \hat p \hat \sigma_z = \hat p |\uparrow \rangle \langle \uparrow |- \hat p|\downarrow \rangle \langle \downarrow |\equiv \hat p_\uparrow - \hat p_\downarrow$ which satisfy the closed set of Heisenberg equations of motion (\ref{Hs1}), with, however, the important difference that $[\hat X, \hat \Pi] = i\hbar \hat \sigma_z$, instead of $[\hat X, \hat \Pi] =0$. The associated Heisenberg uncertainty relation is, therefore, $\Delta \hat X \Delta \hat \Pi \geq  \frac \hbar 2 |\langle \hat \sigma_z \rangle|$.

One might expect that since $\hat X$ and $\hat \Pi$ commute for $\langle \hat \sigma_z\rangle \rightarrow0$ they would form a QMFS in that limit, with an analogous QMFS for the pair of operators $\hat P \equiv \hat p I$ and $\hat \Phi \equiv \hat x \hat\sigma_z$. This is, however, not correct, due to the fact that these are {\it composite} observables of the center-of-mass and spin degrees of freedom. While $\hat X$ has a same expectation value as the center-of-mass position $\hat x$, $\langle \hat \Pi \rangle $ normally differs from $\langle \hat p \rangle $, and can, in particular, be different from zero even for $\langle \hat \sigma_z\rangle = 0$. More importantly, since $\hat \sigma_z^2 = I$ we have $\hat X^2 = \hat \Phi^2$ and  $\hat P^2 = \hat \Pi^2$, so that the separation of the dynamics into two independent dynamical subsystems is invalid for the high-order moments of $\{\hat X, \hat \Pi\}$ and $\{\hat \Phi,\hat P\}$. As a result, measurement backaction, while not affecting the evolution of $\langle\hat X\rangle$ and $\langle\hat\Pi\rangle$ for $\langle \hat \sigma_z\rangle  =0$, does impact their fluctuations, rendering a backaction evading sequence of measurements impossible.

As a concrete example, we consider using a Dirac oscillator operating in the non-relativistic limit  $\epsilon\rightarrow 0$ to perform measurements of a weak spin-dependent external perturbation of the form $V_f=f \hat \sigma_z$ by imprinting its position $\hat x$ on the phase of a quantum harmonic oscillator, the measuring apparatus. The Hamiltonian describing this measurement scheme is  
\begin{equation}
H_{\rm total}= H_{\rm nr}+\hbar \omega_b \hat b^\dagger \hat b + ( g\hat b^\dagger \hat b+ f  \hat \sigma_z  ) \hat x,
\label{eq:V}
\end{equation}
where $b^\dagger$ and $\hat b$ are the creation and annihilation operators of the measuring oscillator, and $g$ a coupling constant. In the non-relativistic limit $d\hat \sigma_z/dt \rightarrow 0$ the Heisenberg equations of motion reduce to
\begin{eqnarray}
\frac{d\hat X}{dt} &=& \hat \sigma_z \frac{\hat p}{m}= \frac{\hat \Pi}{m}, \ \ \ \ \  \frac{d \hat b}{dt} = -i \left( \omega_b + \frac{g}{\hbar} \hat X \right) \hat b\nonumber \\
\frac{d\hat \Pi}{dt} & = &\hat \sigma_z \frac{d\hat p}{dt}  = -m\omega^{2}\hat X - g\hat b^\dagger \hat b \hat \sigma_z  -f, \label{eq:dxdt1}
\end{eqnarray}
where $- g\hat b^\dagger \hat b\hat \sigma_z$ accounts for measurement backaction~\cite{footnote}. 
Solving these equations for $\hat X(t)$ gives
\begin{equation}
\hat X(t)=\hat X(t_0)\cos\omega t+\frac{\hat \Pi(t_0)}{m\omega}\sin\omega t-\frac{2(f+g\hat b^\dagger\hat b\hat\sigma_z)}{m\omega^2}\sin^2\frac{\omega t}{2}.
\label{Xt}
\end{equation}
This confirms that for $\langle \hat \sigma_z\rangle \rightarrow 0$ and initially uncorrelated system and measurement apparatus, $\langle \hat b^\dagger \hat b \hat \sigma_z\rangle =  \langle \hat b^\dagger \hat b \rangle \langle \hat \sigma_z\rangle$,  the measurement of $f$ does not impact the subsequent evolution of $\langle\hat X(t)\rangle$. In particular, for $|\Psi_n (0)\rangle=(|n,\uparrow\rangle+|n-1,\downarrow\rangle)/\sqrt 2$, a superposition that comprises two components of opposite energies and for which $\langle \hat \sigma_z\rangle = 0$, this expression reduces to
\begin{equation}
\langle \hat{X}(t)\rangle=-\frac{2f}{m\omega^{2}}\sin^2\frac{\omega t}{2},
\label{eq:xnr}
\end{equation}
independent of any influence from the measuring apparatus. However its standard deviation is $\Delta\hat X=\sqrt{\langle\hat X^2\rangle-\langle\hat X\rangle^2}=x_{\rm zpt}\sqrt{2n+8G^2\sin^4 (\omega t/2)}$, where $G=\sqrt{2}g\langle\hat b^\dagger\hat b\rangle x_{\rm zpt}/\hbar\omega$ is a dimensionless measurement strength and $x_{\rm zpt} = \sqrt{\hbar/2m\omega}$ is the zero-point width of the oscillator wave function. In addition to the $n$-dependent position uncertainty stemming from the initial state, $\Delta\hat X$ comprises a contribution proportional to $G$, illustrating how measurement backaction limits the precision of subsequent measurements of $f$.

\begin{figure}
\begin{centering}
\includegraphics[scale=0.5]{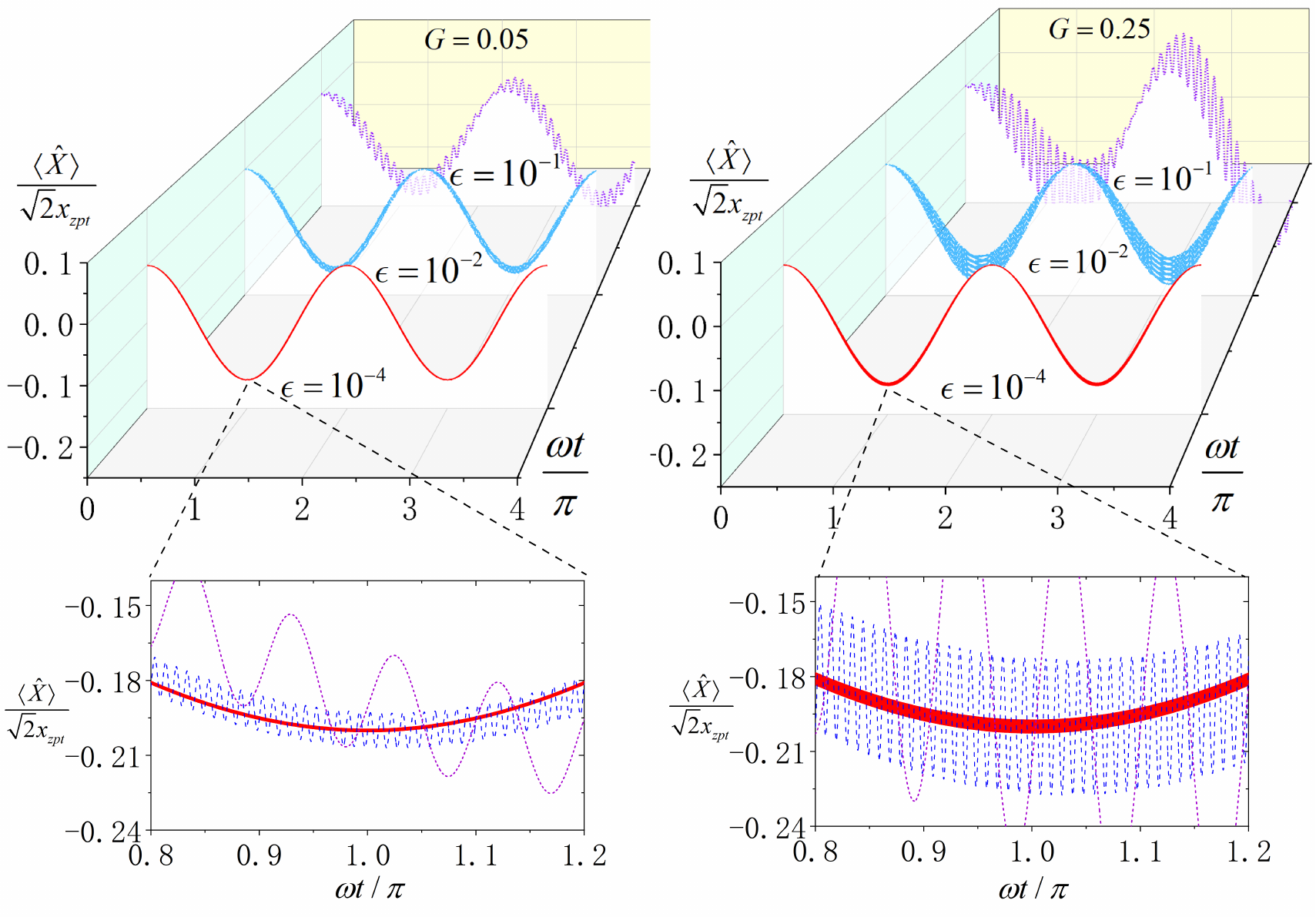}
\par\end{centering}
\caption{Color online: Time evolution of the dimensionless position $\langle\hat X(t) \rangle/\sqrt{2}x_{\rm zpt}$ for a perturbation amplitude $f=0.1$, in units of $\hbar\omega/\sqrt{2}x_{\rm zpt}$, and for increasing relativistic parameters $\epsilon = 10^{-4}$ (red), $10^{-2}$ (blue), and $10^{-1}$ (purple), under the dimensionless measurement strengths $G=\sqrt{2}g\langle\hat b^\dagger\hat b\rangle x_{\rm zpt}/\hbar\omega = 0.05$ (left) and $0.25$ (right). The lower plots show its detailed evolution near the time $\omega t=\pi$. }
\label{error}
\end{figure}

{\it Relativistic backaction} -- Moving past the non-relativistic limit the anharmonicity and spin-orbit coupling of the Dirac oscillator increasingly prevent the conservation of $\langle \hat \sigma_z \rangle$, and measurements disturb not just the higher moments of $\hat X$, but $\langle\hat X(t)\rangle$ as well. Specifically, the spin-orbit coupling generates Zitterbewegung oscillations between the positive and negative energy states of the oscillator, and hence of $\langle\hat\sigma_z\rangle$, at a frequency $\Omega_{\rm zb}$ of the order of $2mc^2/\hbar$.

For small enough $\epsilon$ one can evaluate $\hat \sigma_z(t)$ perturbatively  as $\langle\hat\sigma_z(t)\rangle\approx\sqrt{2n\epsilon}\sin ( 2m c^2 t/\hbar)$, see Supplementary Material, and 
\begin{equation}
\langle \hat{X}(t)\rangle\approx-\frac{2[f + \sqrt{2n\epsilon}g\langle\hat b^\dagger\hat b\rangle \sin ( 2m c^2 t/\hbar) ]}{m\omega^{2}}\sin^2\frac{\omega t}{2}.
\label{corrX}
\end{equation}
It is characterized by fast Zitterbewegung oscillations of small amplitude superposed to slow oscillations at frequency $\omega \ll \Omega_{\rm zb}$ and of amplitude proportional to $2f$, essentially still given by Eq.~(\ref{eq:xnr}). Detecting a signal at the bare frequency of the oscillator results, therefore, in an effective smearing $f\pm\Delta$ of the measured force, with $\Delta=\sqrt{2n\epsilon} g\langle \hat b^\dagger \hat b\rangle$. Since the amplitude of the Zitterbewegung oscillations increases with $G$ the smearing of $f$ can be thought of as a backaction effect that imposes a fundamental limit to the precision of the measurement of $f$~\cite{footnote1}. Alternatively, one can also think of $\Delta$ as an indirect probe of virtual pair creation~\cite{Greiner}.

Measurement backaction increases with $\epsilon$, and the frequency difference between $\omega$ and $\Omega_{\rm zb}$ eventually becomes sufficiently small that $\langle \hat X(t)\rangle$ undergoes anharmonic and aperiodic oscillations.  This is illustrated in Fig.~\ref{error}, which shows numerical simulations of $\langle\hat X(t)\rangle$ for the initial state  $|\Psi_n(0)\rangle $, three values of $\epsilon$ and two values of $G$. 

We note that in the extreme relativistic limit, $\epsilon\gg1$, $H_{\rm DO}$ can be approximated by the Hamiltonian for a Weyl fermion in a spin-dependent potential, $H_{\rm r}=c\hat \sigma_x\hat p-cm\omega\hat \sigma_y \hat x$, in which case spin-orbit coupling fully dominates the dynamics and renders our measurement scheme meaningless.

{\it Implementation} -- The table-top simulation of relativistic quantum systems has witnessed considerable recent progress. The use of a single trapped ion to simulate the Dirac equation was proposed by Lamata {\it et al.}~\cite{Lamata2007} and experimentally realized by Gerristma {\it et al.}~\cite{Gerritsma2010}. Solano and coworkers~\cite{Solano2007} demonstrated the mapping of the Dirac oscillator onto the Jaynes-Cummings model. Reference~\cite{Seligman2013} reported the realization of a Dirac oscillator in a microwave system, with further proposals involving fiber Bragg gratings~\cite{Longhi2010}, hexagonal structures of graphene~\cite{Mortessagne2010}, superconducting qubits~\cite{Solano2013}, and optomechanical arrays~\cite{Marquardt2015}.

\begin{table*}
\begin{tabular}{|c | c | c| c| c|}
\toprule
Dirac oscillator & Electron & SOC condensate & cQED & Ion\\
\hline
\midrule
Velocity of light (m/s) & $c\sim10^8$ & $\tilde c=\hbar k_r/m_a\sim10^{-2}$ & $\sim 10$ & $\sim 10^{-3}$\\
\hline
Rest mass (kg) & $m_e\sim10^{-31}$ & $\tilde m=\chi m_a^2/\hbar k_r^2\sim10^{-27}$ & $\sim10^{-26}$ &$\sim10^{-23}$\\
\hline
Reduced Compton wavelength (m)& $\lambdabar_c=\hbar/m_e c\sim 10^{-12}$ & $\tilde\lambdabar_c=\hbar k_r/\chi m_a\sim10^{-5}$ & $\sim10^{-9}$ & $\sim 10^{-8}$\\
\hline
Zitterbewegung frequency ($2\pi\times$Hz)& $\Omega_{zb}=2m_e c^2/\hbar\sim 10^{21}$ & $\tilde\Omega_{zb}=2\chi\sim 10^3$ & $\sim 10^{10}$ &$\sim 10^5$\\
\hline
Oscillator frequency ($2\pi\times$Hz)& $\omega$ & $\tilde\omega=\hbar k_r\varsigma/m_a\chi\sim 0-10^4$ & $\sim 10^{10}$ & $\sim 10^6$\\
\hline
Relativistic parameter & $ \epsilon=\hbar\omega/m_e c^2$ & $\tilde\epsilon=\hbar k_r \varsigma/m_a\chi^2\sim0-10$ & $\sim 1$ &$\sim 10$\\
\bottomrule
\end{tabular}
\caption{Mapping of key physical quantities of the original Dirac oscillator onto its spin-orbit-coupled condensate implementation, with magnitudes taken from the experiments of Refs.~\cite{Spielman2013} and \cite{Engels2014}. The magnitude of the corresponding parameters for possible circuit-QED and ion implementations are also given for comparison, based on Refs.~\cite{Blais2004} and \cite{Lamata2007}. }
\label{table}
\end{table*}

We propose here an alternative scheme based on a spin-orbit coupled (SOC) atomic condensate~\cite{Spielman2011}, as its macroscopic coherence provides considerable flexibility toward the realization of measurements spanning a broad parameter range from the effective `non-relativistic' to  a `strongly relativistic' limit. We consider an atomic Bose-Einstein condensate (BEC) of three-level atoms with a pair of pseudo-spin hyperfine lower states optically coupled by two Raman fields far-off resonant from an upper electronic state that is adiabatically eliminated from the dynamics. The resulting transitions between hyperfine states are accompanied by a momentum transfer of $2\hbar k_r$ in the $x$ direction, where $\hbar k_r$ is the photon recoil momentum. The dynamics in the other two directions is assumed to be decoupled, resulting in an effective one-dimensional situation. 

In the mean-field approximation the dynamics of the condensate can be described by a 1D Gross-Pitaevskii equation for the Hamiltonian $H_{\rm s}+H_{\rm c}$, where $H_{\rm s}$ is the single-atom Hamiltonian and $H_{\rm c}$ accounts for two-body collisions. In momentum representation and the basis of the two hyperfine states $H_{\rm s}$ is
\begin{equation}
H_{\rm s}=\begin{pmatrix}\frac{\hbar^{2}(k+k_{r})^{2}}{2m_a}+\frac{\hbar\delta}{2} & \hbar\Omega\\
\hbar\Omega^{*} & \frac{\hbar^{2}(k-k_{r})^{2}}{2m_a}-\frac{\hbar\delta}{2}
\end{pmatrix},
\label{eq:Hs}
\end{equation}
where $m_a$ is the atomic mass and $\delta$ and $\Omega$ are the two-photon detuning and effective Rabi coupling. This is similar to the model describing recent experiments on SOC BEC \cite{Pan2012, Spielman2013, Engels2014, Ketterle2017}, with the difference that we assume that the real part of the two-photon Rabi coupling $\Omega$ has a linear spatial dependence, $\Omega=|\Omega| e^{i\phi(x)}$,  with $\phi(x)$ slowly varying over the length of the condensate and $\phi(x_0) = \pi/2$ at its center $x_0$ so that $\Omega(x_0)$ is purely imaginary. This can be achieved by controlling the relative phase of the two Raman fields with a spatial phase modulator. We then have
\begin{equation}
\Omega \approx - \hat x |\Omega(x_0)| \frac{\partial \phi(x)}{\partial x} \bigg\rvert_{x_0} +\Omega(x_0)\equiv -\varsigma \hat x + i\chi
\label{eq:omega}
\end{equation}
where $\chi = {\rm Im}[\Omega(x_0)]$. Hence the real part of the two-photon Rabi frequency has a linear dependence on position, while its imaginary part is constant. With the matrix representation of Pauli operators and  after a pseudo-spin rotation $\sigma_{x}\rightarrow\sigma_{y}\rightarrow\sigma_{z}\rightarrow\sigma_{x}$,  Eqs.~(\ref{eq:Hs}) and (\ref{eq:omega}) give, for $\delta = 0$ and $k\ll k_{r}$, 
\begin{equation}
H_s=\frac{\hbar^2 k_r ^2}{2m_a}+\frac{\hbar^{2}k_r}{m_a}k\hat \sigma_x-\hbar\varsigma\hat x\hat\sigma_y+\hbar\chi\hat{\sigma}_z.
\label{eq:hs}
\end{equation}

Except for the kinetic energy term this Hamiltonian has the same form as $H_{\rm DO}$, with effective mappings between the velocity of light $c$ and the atomic recoil velocity $\hbar k_r/m$, and between the rest energy $mc^2$ and the Rabi coupling energy $\hbar\chi$. Here $k_{r}$, $\varsigma$, and $\chi>0$. Importantly, the kinetic energy term raises up the zero energy level to $\hbar^{2}k_{r}^{2}/2m_a$, allowing the physical implementation of an analog of negative-energy states. See Table \ref{table} for the full mapping between the two systems and an estimate of the order of magnitude of the key parameters.

The two-body collisions are described by the Hamiltonian
\begin{equation}
H_{\rm c}=\begin{pmatrix}g_{\uparrow\uparrow}\left|\psi_{\uparrow}\right|^{2}+g_{\uparrow\downarrow}\left|\psi_{\downarrow}\right|^{2} & 0\\
0 & g_{\uparrow\downarrow}\left|\psi_{\uparrow}\right|^{2}+g_{\downarrow\downarrow}\left|\psi_{\downarrow}\right|^{2}
\end{pmatrix},
\end{equation}
where $\psi_{\uparrow,\downarrow}$ are the many-body wave functions for atoms in the spin-up and spin-down hyperfine levels, and $g_{ij}=4\pi \hbar ^2 a_{ij}/m$, with $a_{ij}$ the corresponding scattering lengths, measure the effective inter- and intra-spin collision strengths. When collisions dominate the condensate can be described in a single-mode approximation, as is the case e.g. for $^{87}{\rm Rb}$, for which $g_{\uparrow \uparrow} =  g_{\uparrow \downarrow} \approx g_{\downarrow \downarrow}$ \cite{Engels2014}. This results in the effective Hamiltonian
\begin{equation}
\mathcal{H}=\frac{2\tilde c}{\hbar}\hat{S}_{x}\hat{p}-\frac{2\tilde c \tilde m \tilde\omega}{\hbar}\hat{S}_{y}\hat{x}+\frac{2\tilde m \tilde c^{2}}{\hbar}\hat{S}_{z},
\end{equation}
where $\hat S_{x,y,z}=\hbar/2\sum_i^N\hat{\sigma}_{x,y,z}^i$ are collective spin operators with $N$ the number of atoms, and we have neglected terms proportional to  $g_{\uparrow\uparrow}-g_{\downarrow\downarrow}$ and $g_{\uparrow\uparrow}+g_{\downarrow\downarrow}-2g_{\uparrow\downarrow}$. This Hamiltonian has the same form as $H_{\rm DO}$, with the substitution of $\hat \sigma_{x,y,z}$
by $\hat S_{x,y,z}$.

The non-relativistic limit of $\mathcal H$ is approached under the strong Raman coupling condition $\chi\gg\sqrt{\hbar k_{r}\varsigma/m_a}$, resulting in an approximate Hamiltonian 
\begin{equation}
\mathcal{H}_{\rm nr}=\left (2\tilde m \tilde c^{2}+\frac{\hat p^2}{\tilde m}+\tilde m \tilde\omega^{2}\hat{x}^{2}\right )\hat S_z/\hbar. 
\end{equation}
which has the same form as Eq.~(\ref{eq:nrDirac}), so that the previous discussion can be readily applied~\cite{footnote2}. 

For a measurement interaction of the form 
\begin{equation}
\mathcal{V}_{\rm nr}=Ng\hat{b}^{\dagger}\hat{b}\hat{x}+2f\hat{S}_{z}\hat{x}/\hbar
\end{equation} 
the coupling to the perturbation $f$, of the form $2f\hat S_y \hat x/\hbar $ in the original physical representation, could be realized through a spatially dependent Raman coupling such that the spatial dependence appears now in the imaginary part of the effective Rabi frequency. The interaction with the measurement apparatus, $Ng\hat x\hat b^\dagger\hat b$, can be realized by an optomechanical-like collective coupling between the condensate and a cavity field $\hat b$ \cite{Esslinger2008, Stamper-Kurn2008}. As with the original Dirac oscillator, the departure from the non-relativistic regime resulting from a decrease in $\hbar\chi$ results in an increase in backaction, now from the analog of Zitterbewegung oscillations~\cite{Clark2008, Spielman2013}. 

In that limit $\mathcal H$ simplifies to
\begin{equation}
\mathcal{H}_{\rm r}=\frac{2\tilde c}{\hbar}\hat{S}_{x}\hat{p}-\frac{2\tilde c \tilde m \tilde\omega}{\hbar}\hat{S}_{y}\hat{x},
\end{equation}
compare to the Weyl fermion Hamiltonian $H_{\rm r}$. For a sufficiently large number of atoms highly polarized along the direction of $\hat{S}_{z}$, a Holstein-Primakoff transformation is sometimes invoked to map the collective spin operators to position and momentum operators of an effective oscillator, $\hat{x}_{s}=\hat{S}_{y}/\sqrt{|S_{z}|}$, $\hat{p}_{s}=-\text{sgn}(S_{z})\hat{S}_{x}/\sqrt{|S_{z}|}$, and $S_{z}\approx\pm\hbar N/2$ \cite{Polzik2017} so that 
\begin{equation}
\mathcal{H}_{\rm r} \approx -\tilde c\sqrt{2N/\hbar}\left [ {\rm sgn }(S_z) \hat p \hat p_s + \tilde m\tilde\omega \hat x \hat x_s\right ],
\end{equation}
with $[\hat x, \hat p_s] = 0$. For the coupling $\mathcal{V}_{\rm r}=Ng\hat{b}^{\dagger}\hat{b}\hat{x}+2f\hat{S}_{y}/\hbar
$, $\hat x$ and $\hat p_s$ now appear to constitute  a true QMFS. However, this relies on neglecting the quantum fluctuations of $\hat S_z$, that is, on treating it as a classical quantity. (The same would hold for $H_{\rm r}$ if $\hat \sigma_z$ was treated classically.)  But the fundamental reason why a QMFS cannot be realized in the Dirac oscillator is precisely that $\sigma_z$ is an operator, and that same issue appears in the atomic system as well. In that case, measurement backaction would serve as a probe of the limitations of the Holstein-Primakoff approximation.

{\it Summary and outlook} -- Summarizing, we have shown that despite the existence of negative energy states the Dirac oscillator is fundamentally different from a system of two harmonic oscillators with equal and opposite masses, and as a consequence can not operate as a QMFS back-action evading detector. The origin of this difference is a spin-orbit coupling interaction that results in the relativistic regime in Zitterbewegung oscillations. We suggested that measurement backaction can be exploited as a probe of the associated virtual pair creation, and proposed a tabletop demonstration of this effect in a spin-orbit-coupled atomic condensate.

\begin{acknowledgments}
We acknowledge enlightening discussions with J. Chen and W. Xie. Work was supported by the National Key Research and Development Program of China (Grant No. 2016YFA0302001), the National Natural Science Foundation of China (Grant Nos. 11574086, 91436211, 11654005, 11234003, and 11374003), the Shanghai Rising-Star Program (Grant No. 16QA1401600), and the Science and Technology Commission of Shanghai Municipality (Grant No. 16DZ2260200).
\end{acknowledgments}

%%%%%%%%%% Merge with supplemental materials %%%%%%%%%%
%%%%%%%%%% Prefix a "S" to all equations, figures, tables and reset the counter %%%%%%%%%%
\setcounter{equation}{0}
\setcounter{figure}{0}
\setcounter{table}{0}
\setcounter{page}{1}
\makeatletter
\renewcommand{\theequation}{S\arabic{equation}}
\renewcommand{\thefigure}{S\arabic{figure}}
\renewcommand{\bibnumfmt}[1]{[S#1]}
\renewcommand{\citenumfont}[1]{S#1}
%%%%%%%%%% Prefix a "S" to all equations, figures, tables and reset the counter %%%%%%%%%%

\section*{Supplementary material}

\subsection{Estimation of the relativistic backactions}

The original Hamiltonian $H'_{\rm total}=H_{\rm DO}+\hbar\omega_b \hat b^\dagger \hat b+V$ can be written in the form of the anti-Jaynes-Cummings Hamiltonain
\begin{eqnarray}
H'_{\rm total}&=&mc^2\hat\sigma_z+(ic\sqrt{\frac{\hbar\omega m}{2}}\hat a^\dagger\hat\sigma_+ + {\rm h.c.})\nonumber\\
&&+\hbar\omega \hat b^\dagger \hat b+\sqrt{\frac{\hbar}{2m\omega}}(g\hat b^\dagger \hat b+f\hat\sigma_z)(\hat a+\hat a^\dagger),
\end{eqnarray}
where 
\begin{equation}
\hat a=\hat x\sqrt{m\omega/2\hbar}+i\hat p/\sqrt{2m\hbar\omega}
\end{equation}
is a motional bosonic annihilation operator and $\hat \sigma_+=\hat \sigma_x+i\hat\sigma_y$. The corresponding Heisenberg equations of motion  are
\begin{eqnarray}
i\hbar\frac{d\hat\sigma_+}{dt}&=&-2\left [ mc^2+\sqrt{\frac{\hbar}{2m\omega}}f(\hat a+\hat a^\dagger) \right ]\hat\sigma_+ -4ic\sqrt{\frac{\hbar\omega m}{2}}\hat\sigma_z\hat a. \nonumber\\
\label{dsigma+}\\
i\hbar\frac{d\hat \sigma_z}{dt}&=&2ic\sqrt{\frac{\hbar\omega m}{2}}\hat a^\dagger\hat\sigma_+-{\rm h.c.}\label{dsigmaz}\\
i\hbar\frac{d\hat a}{dt}&=&ic\sqrt{\frac{\hbar\omega m}{2}}\hat\sigma_+ +\sqrt{\frac{\hbar}{2m\omega}}(g\hat b^\dagger \hat b+f\hat\sigma_z) \label{da}
\end{eqnarray}
The oscillations of $\hat \sigma_\pm$ occur at a frequency of the order $2mc^2/\hbar$, which results in Zitterbewegung oscillations between the positive- and negative-energy states. For $mc^2/\hbar\gg\omega$, we can adiabatically eliminate the dynamical equation of $\hat\sigma_+$ by inserting the approximate expression
\begin{equation}
\hat\sigma_+\approx \left[ \frac{-i\sqrt{2\hbar\omega/mc^2}}{1+\frac{f}{mc^2}\sqrt{\frac{\hbar}{2m\omega}}(\hat a+\hat a^\dagger)}\right]\hat\sigma_z\hat a
\end{equation}
obtained from Eq. (\ref{dsigma+}) into Eqs.~(\ref{dsigmaz}) and (\ref{da}). They reduce then to
\begin{eqnarray}
i\hbar\frac{d\hat\sigma_z}{dt}&\approx& \hat a^\dagger \left[ \frac{2\hbar\omega}{1+\frac{f}{mc^2}\sqrt{\frac{\hbar}{2m\omega}}(\hat a+\hat a^\dagger)} \right] \hat a\hat\sigma_z-{\rm h.c.}=0, \nonumber \\
\frac{d\hat a}{dt}&\approx&-i\left[ \frac{\omega}{1+\frac{f}{mc^2}\sqrt{\frac{\hbar}{2m\omega}}(\hat a+\hat a^\dagger)} \right]\hat\sigma_z\hat a\nonumber \\
&&-\frac{i}{\sqrt{2\hbar\omega m}}(g\hat b^\dagger\hat b+f\hat\sigma_z).
\end{eqnarray} 
For a weak perturbation $f\langle\hat x\rangle \ll mc^2$ we can ignore the operator part in the denominator, and with the definition of $\hat a$ we then obtain the  approximate Heisenberg equations of motion
\begin{eqnarray}
\frac{d\hat x}{dt} &=& \frac{\hat\sigma_z}{m}\hat p, \\
\hat\sigma_z\frac{d \hat p}{dt} &=& -m\omega^{2}\hat x - g\hat b^\dagger \hat b \hat \sigma_z - f.
\end{eqnarray}
Their solution is
\begin{eqnarray}
\hat x(t)&=&\hat x(t_0)\cos\omega t+\frac{\hat p(t_0)\hat\sigma_z}{m\omega}\sin\omega t-\frac{2(f+g\hat b^\dagger\hat b\hat\sigma_z)}{m\omega^2}\sin^2\frac{\omega t}{2},\nonumber\\
\label{xt}\\
\hat p(t)&=&\hat p(t_0)\cos\omega t-\hat x(t_0) \hat\sigma_z m\omega\sin\omega t-\frac{f\hat\sigma_z+g\hat b^\dagger\hat b}{\omega}\sin\omega t.\nonumber\\
\label{qt}
\end{eqnarray} 
For a spin balanced initial state such as 
\begin{equation}
|\Psi_n (t_{0})\rangle=\frac{| n,\uparrow\rangle+|n-1,\downarrow\rangle}{\sqrt 2}.
\end{equation}
For $\langle\hat\sigma_z\rangle=0$, the evolution of $\hat x(t)$ is independent on the measuring apparatus $\hat b$.

As $\epsilon$ increases the approximation $d\hat\sigma_z/dt\approx 0$ gradually loses its validity and the measurement backaction term, $g\hat b^\dagger\hat b\hat\sigma_z$, can no longer be neglected in the evolution of $\hat x$. When the measurement interaction is much weaker than the original energy of the oscillator, $V\ll mc^2$, we can estimate the evolution of $\langle\hat\sigma_z(t)\rangle$ approximately by the Heisenberg equations with Hamiltonian $H_{\rm DO}$ only. After expanding the initial state $|\Psi_n (t_{0})\rangle$ on the eigenstates of $H_{\rm DO}$ we can obtain the evolution of the state, and hence $\langle \hat \sigma_z \rangle$ as
\begin{equation}
\langle\hat\sigma_z(t)\rangle\approx\sqrt{\frac{2n\epsilon}{1+2n\epsilon}}\cos(\frac{2E_n^+t}{\hbar})\overset{\epsilon\ll 1}{\approx}\sqrt{2n\epsilon}\sin(\frac{2mc^2 t}{\hbar})
\end{equation} 
from which we can estimate the influence of the measurement backaction on the oscillations of $\hat X(t) \equiv\hat x I$ from Eq.~(\ref{xt}). This gives
\begin{equation}
\langle \hat{X}(t)\rangle\approx-\frac{2[f+\sqrt{2n\epsilon}g\langle\hat b^\dagger\hat b\rangle\sin(\frac{2mc^2 t}{\hbar}) ]}{m\omega^{2}}\sin^2\frac{\omega t}{2},
\label{corr1}
\end{equation}
which shows that the backaction error increases with the measurement strength $g\langle\hat b^\dagger\hat b\rangle$.

\subsection{Measurement in the Foldy\---Wouthuysen representation }

The complications associated with the role of spin-orbit coupling can be formally eliminated in the Foldy\textendash Wouthuysen (FW) representation~\cite{Moreno1989, Toyama1997}, where the positive- and negative-energy states of the Dirac oscillator are completely separated by the different spinors, as the case of the non-relativistic limit. The FW and the original Dirac representations are related by the nonlocal transformation $\hat U=e^{\beta\frac{\alpha\cdotp\hat{\pi}}{2|\hat{\pi}|}\theta}$, with $\theta=\arctan(|\hat{\pi}|/mc)$ and $\hat{\pi}=\hat{p}-i\beta m\omega\hat{x}$.

In the FW representation $H_{\rm DO}$ is diagonalized into the form
\begin{equation}
H_{\rm FW}=\hat UH_{\rm DO}\hat U^{\dagger}=\tilde{\sigma}_{z}c\sqrt{(mc)^{2}+\tilde{p}^{2}+(m\omega\tilde{x})^{2}-\tilde{\sigma}_{z}m\hbar\omega},
\end{equation}
where $\tilde{x}$, $\tilde{p}$, and $\tilde{\sigma}_z$ are the position, momentum, and spin operators in the FW representation.
The corresponding eigenenergies are still $E_n^\pm$ as in the main text, and the corresponding eigenstates, given by the transformation 
\begin{eqnarray}
|E_n^+ \rangle_{\rm FW}&=&\hat U |E_n^+ \rangle=|n,\uparrow \rangle,\nonumber\\
|E_n^- \rangle_{\rm FW}&=&\hat U |E_n^- \rangle=|n,\downarrow \rangle,
\end{eqnarray}
are same as the reduced eigenstates of Dirac representation in the non-relativistic limit.  For the total Hamiltonian $H_{\rm FW}+V$, $d\tilde\sigma_z/dt$ remains exactly zero at all times, but due to the anharmonic form of $H_{\rm FW}$, the dynamics also involves higher moments through the commutators $[\tilde{x},H_{\rm FW}^{n}]=i\hbar nc^{2}H_{\rm FW}^{n-2}\tilde{p}$ and $[\tilde{p},H_{\rm FW}^{n}]=-i\hbar n(m\omega c)^{2}H_{\rm FW}^{n-2}\tilde{x}$ $(n\geq 1)$, so that the composite operators $\tilde X \equiv \tilde x \boldsymbol {\it I}$ and $\tilde \Pi \equiv \tilde p \tilde \sigma_z$ can not constitute a QMFS. 

If the measurement interaction is much weaker than the original energy of the oscillator, that is, $|V/H_{\rm FW}|\ll1$ we can obtain the Heisenberg evolution of $\tilde{x}$ through the perturbation theory,
\begin{align}
\tilde x(t) & =\mathcal{T}^{\dagger}(t,t_0)\tilde x\mathcal{T}(t,t_0)\nonumber \\
 & +\frac{1}{i\hbar}\int_{0}^{t}dt^{\prime}[\mathcal{T}^{\dagger}(t,t_0)\tilde x\mathcal{T}(t,t_0),\mathcal{T}^{\dagger}(t,t^{\prime})V\mathcal{T}(t,t^{\prime})]+O(V^{2}),
\end{align}
where the time evolution operator $\mathcal{T}(t,t^{\prime})=e^{-iH_{\rm FW}(t-t^{\prime})/\hbar}$. The resulting approximate evolution of $\hat x$ takes a form similar to Eq.~(16) in the main text, 
\begin{eqnarray}
\tilde{x}(t) &\approx &\tilde{x}(t_0)\cos(\hat\omega t)\nonumber\\
&+&\frac{\tilde{p}(t_0)\tilde{\sigma}_{z}}{m\omega}\sin(\hat\omega t)-\frac{2(g\tilde{\sigma}_{z}\hat{b}^{\dagger}\hat{b}+f)}{m\omega\hat\omega}\sin^{2}(\frac{\hat\omega t}{2}),\label{xxt}
\end{eqnarray}
but with an important difference that the effective oscillation frequency $\hat\omega=\tilde{\sigma}_{z}m\omega c^{2}/H_{\rm FW}$ is in an operator form.

Measurement backaction disappears in the evolution of $\langle\tilde{x}\rangle$ therefore when the DO is initially in an `energy balanced' state such that $\langle H_{FW}\rangle=0$, a more stringent condition than in the case of non-relativistic limit where only $\langle\hat\sigma_z\rangle=0$ is required. For instance when the initial state is $|\Psi(t_{0})\rangle=(|E_n ^+\rangle_{\rm FW} +|E_{n-1}^-\rangle_{\rm FW} )/\sqrt{2}$, Eq. (\ref{xxt}) will give a position evolution
\begin{equation}
\left\langle \tilde{x}(t)\right\rangle =-\frac{2E_{n}^{+}f}{(m\omega c)^{2}}\sin\left ( \frac{mc^2\omega t}{2E_n^+ }\right ),
\end{equation}
which is independent on the measuring apparatus $\hat b$. Note that when $mc^{2}\gg\hbar\omega$ we have $E^+_{n}\approx mc^{2}$, so this result is consistent with the result obtained in the non-relativistic limit ( Eq.~(18) in the main text ). Similarly, the influence of measuring apparatus appears in the fluctuation of $\tilde{x}$, so the measurement backactions is unavoidable as in the physical Dirac representation.

Although Zitterbewegung oscillations for a general DO appears to be avoidable in the FW representation, the problem is that when returning to the physical Dirac representation, $\tilde{x}$ becomes the Newton-Wigner (NW) position operator, $\hat X_{\rm NW}=\hat U^{\dagger}\tilde{x}\hat U$, which is not diagonal in coordinate space, accompanied by a positional smearing over a region comparable to the Compton wavelength of the particle. Measuring $\hat{x}_{\rm NW}$ might require a full quantum tomography of the evolved state of the system or require some additional Ramsey-like rotations before measurements~\cite{Solano2013}. Hence, a precise measurement of NW position does not imply that this is the case for the physical position. Instead, a fundamental uncertainty of size $\sim\hbar/mc$ is inevitable. Besides, it also requires a more special form of the measurement interaction than in the Dirac representation, $V_{\rm NW}=\hat U^{\dagger}V\hat U$.

The explicit forms of $\hat{X}_{\rm NW}$ and $V_{\rm NW}$ in the Dirac representation are too complicated to reproduce here, but we can obtain approximate expressions via a series expansion according to the order of the ratio $\hbar\omega/mc^{2}$. For instance, the leading-order approximation gives exactly the non-relativistic limit results. In the first-order approximation the expression of the NW position operator is
\begin{equation}
\hat{X}_{\rm NW}\approx\hat{x}-\frac{\hbar}{2mc}\hat{\sigma}_{y},
\end{equation}
and the measurement interaction is
\begin{equation}
V_{\rm NW}\approx V-\frac{\hbar g}{2mc}\hat{b}^{\dagger}\hat{b}\hat{\sigma}_{y}+\frac{f}{2mc}(\hat{x}\hat{p}+\hat{p}\hat{x})\hat{\sigma}_{x}.
\end{equation}
which shows that the coupling to the perturbation $f$ to be measured is in a complex spin-orbit-coupled form.


\begin{thebibliography}{99}
\bibitem{footnote} If the external perturbation $f$ is spin independent then for a spin-dependent phase imprinting, we can obtain similar dynamical equations for $\hat\Phi$ and $\hat P$.
\bibitem{footnote2} Remember, however, that a pseudo-spin rotation was applied in the derivation of $H_{\rm s}$, see Eq.~(\ref{eq:hs}); this means that the initial atomic states must satisfy the condition $\langle\hat{S}_{y}\rangle=0$ in the original physical representation.
\end{thebibliography}
\end{document}